\documentclass[]{pasj01}

\usepackage{graphicx}	

\newcommand{\Modot}{{M_{\rm \odot}}}

\def\Msun{$M_{\rm \odot}$}

\def\arcdeg{$^\circ$}

\def\kmskpc{\,km\,s$^{-1}$\,kps$^{-1}$}

\Received{}
\Accepted{}
 
\begin{document} 


\title{ 
Dynamical Influence of a Central Massive Object on Double-Barred Galaxies: Self-Destruction Mechanism of Secondary Bars
}

\author{Naoki \textsc{Nakatsuno}, \altaffilmark{1}}%

\author{Junichi \textsc{Baba}\altaffilmark{2,3}}
\email{babajn2000@gmail.com; junichi.baba@sci.kagoshima-u.ac.jp}

\altaffiltext{1}{Department of Astronomy, Graduate School of Science, The University of Tokyo, 7-3-1 Hongo, Bunkyo-ku, Tokyo 113-0033, Japan.}
\altaffiltext{2}{Amanogawa Galaxy Astronomy Research Center, Graduate School of Science and Engineering, Kagoshima University, 1-21-35 Korimoto, Kagoshima 890-0065, Japan.}
\altaffiltext{3}{National Astronomical Observatory of Japan, Mitaka, Tokyo 181-8588, Japan.}


\maketitle

\begin{abstract}
Double-barred galaxies exhibit sub-kpc secondary stellar bars that are crucial for channeling gases towards a central massive object (CMO) such as a supermassive black hole or a nuclear star cluster. Recent $N$-body simulations have uncovered a novel galaxy evolution scenario wherein the mass of the CMO increases owing to the secondary bar, resulting in the eventual destruction of the latter. Consequently, the CMO mass growth halts, thus suggesting a maximum CMO mass of $\approx 10^{-3}$ of the stellar mass of the galaxy. 
This study focused on backbone orbit families, particularly double-frequency orbits, within double-barred galaxies. Consequently, the dynamic influence of a CMO on these orbits was investigated.
The results of the study revealed the emergence of a new orbital resonance within the central region of the galaxy upon the introduction of a CMO. Orbits subjected to this resonance become chaotic and fail to support the secondary bar, ultimately resulting in the destruction of the entire structure. This is partly because of the inability of the secondary bar to obtain support from the newly generated orbit families following the appearance of resonance. Through the estimation of the condition of secondary bar destruction in realistic double-bar galaxies with varying pattern speeds, the results of the study established that such destruction occurred when the CMO mass reached $\approx 10^{-3}$ of the galaxy mass.
Furthermore, a physical explanation of the galaxy evolution scenario was provided, thereby elucidating the interaction between the CMO and the secondary bar. The understanding of the co-evolution of the secondary bar and the CMO, based on stellar orbital motion, is a crucial step towards future observational studies of stars within the bulge of the Milky Way.
\end{abstract}

\section{Introduction}

Numerous galaxies contain a supermassive black hole (SMBH) and/or nuclear star cluster (NSC) at their core. The instances of co-existing SMBHs and NSCs within a galaxy have been extensively documented \citep{Seth+2008}.
The mass of an SMBH is typically in the range $10^6$--$10^9~\Modot$. Further, the masses exhibit correlations with the properties of the elliptical or classical bulges of their host galaxies \citep{KormendyHo2013}. NSCs, which are compact stellar systems with effective radii of approximately 3 pc and masses of approximately $10^5$--$10^8~\Modot$, also display scaling relationships with their host galaxies \citep{Neumayer+2020}. 
The presence of these scaling relations between ``central massive objects'' (CMOs).

Galaxy mergers are hypothesized to generate the observed CMO--host galaxy correlations \citep{DiMatteo+2008,Antonini+2012}. However, disc galaxies with pseudo-bulges, such as the Milky Way, which are predominantly formed via merger-free processes, also harbor SMBHs and NSCs \citep{Greene+2020,Neumayer+2020}. This finding implies that substantial CMO growth may be driven primarily by merger-free internal (secular) processes \citep{Simmons+2017,Martin+2018}.
This study focused on the mass assembly mechanism of CMOs owing to gas inflow within a disc galaxy.

Non-axisymmetric galactic structures, such as bars and spiral arms, possess kpc-scale dimensions and alter the angular momentum of the gas in a galactic disc. This results in gas inflow into the central sub-kpc regions of the galaxies \citep{Athanassoula1992b,ReganTeuben2003,Li+2015,Sormani+2018b,BabaKawata2020a,Tress+2020}.
However, the gas reaches the central sub-kpc regions and forms nuclear rings approximately 100 pc from the center because of the decreased efficiency of the large-scale torques attributed to bars/spirals. Within the 10 pc scale surrounding the CMO, various angular momentum transfer mechanisms that are not related to galactic gravity fields have been proposed. These include radiative drag from the accretion disc around the SMBH or supernova-driven turbulence in a circumnuclear disc \citep{Wada+2002,KawakatuUmemura2002,Izumi+2016}. Consequently, numerous models have introduced non-axisymmetric gravitational structures to facilitate the continuous transportation of gas to smaller radii \citep{HopkinsQuataert2010}. 

A short secondary bar, with a general radius of less than 1 kpc, is a pervasive structure observed in 25--40\% of local barred galaxies \citep{ErwinSparke2002,Laine+2002}. Such systems are known as double- or nested-barred systems.
Numerical simulations suggest that short secondary bars are induced by various instabilities in the central sub-kpc regions of barred galaxies, with secondary and primary bars dynamically decoupled from each other \citep{FriedliMartinet1993,Heller+2001,Rautiainen+2002,DebattistaShen2007,Wozniak2015}.
In the potential of two independently rotating bars, there exists a supporting orbit family whose self-consistency with the underlying double-barred potentials primarily is dependent on the ratio of the pattern speeds of the primary and secondary bars \citep{MaciejewskiSparke2000,MaciejewskiAthanassoula2008}. 
In this context, a short secondary bar is a promising candidate for driving the gas inflow into the center and ensuring efficient feeding of the CMOs \citep{Shlosman+1989,EnglmaierShlosman2004,Namekata+2009}. 
There is observational evidence of gas inflow in double-barred galaxies; for example, in NGC6946 \citep{Schinnerer+2006,Schinnerer+2007}.
Nevertheless, CMO growth owing to gas inflow driven by the secondary bar can affect the supporting orbits of the secondary bar itself.

\citet{Du+2017} conducted a pioneering study that investigated the co-evolution of the short secondary bar and CMO through self-consistent $N$-body simulations. They executed $N$-body simulations of disc galaxy models composed of a stellar disc embedded in a dark matter halo. The aim was to spontaneously generate a double-bar structure \citep{Du+2015}. 
Subsequently, they introduced a CMO mass increase attributed to the gas inflow. They posited that the secondary bar was destroyed when the CMO mass ($M_{\rm CMO}$) reached $\approx 5\times10^{-5}$--$2\times10^{-3}$
of the galaxy stellar mass ($M_{\ast}$).
This implies the existence of a ``self-destruct'' mechanism wherein the secondary bar transports gas and promotes CMO mass gain. However, when $M_{\rm CMO}/M_\ast$ reaches $\approx 10^{-3}$, the destruction of the secondary bar stops the gas supply and consequently, the CMO's mass gain. 
\citet{Du+2017} suggested that such a self-destruct scenario can explain the observed relationship between the lack of the secondary bar and $M_{\rm CMO}/M_\ast \approx {10^{-3}}$.

Furthermore, \citet{Guo+2020} analyzed their $N$-body simulations to examine the dynamic evolution of a secondary bar destroyed by CMO growth. The results demonstrated that the secondary bar can transform into a classical bulge. This presents a novel channel for classical bulge formation, as classical bulges are assumed to be formed via violent dynamical effects such as galaxy mergers \citep{Toomre1977merger,BrooksChristensen2016} or massive clump instabilities in high-z disks \citep{Bournaud2016}. Consequently, a new formation pathway was proposed, wherein classical bulges were formed through the destruction of the secondary bar. However, the physical mechanisms of secondary bars' self-destruction and subsequent transformation into classical bulges remain unclear.

In this study, we aimed to provide a physical interpretation of the dynamic effects of CMOs on orbits within a double-barred galaxy. We focused on the backbone orbits supporting a double-barred galaxy and explored changes in the nature of these supporting orbits with the mass of the CMO. These supporting orbits were multiperiodic orbits, referred to as loops, which were proposed by \citet{MaciejewskiSparke1997,MaciejewskiSparke2000}.
We emphasize that this study is not aimed at constructing a self-consistent model but rather to investigate the changes in the parent orbits (i.e., stable loops) of a double-barred galaxy with variation in the CMO mass.

The remainder of this paper is organized as follows. Section \ref{sec:ModelMethod} describes the galaxy model and analysis methods for examining the orbital properties. 
Section \ref{sec:RESULTS} presents the results. The CMO growth was found to generate new orbital resonances in the secondary bar region, which significantly increased the chaos in the supporting orbits of the secondary bar. This result extends the results of a previous study on supporting orbits in double bars \citep{MaciejewskiSparke2000}, and provides a physical explanation for the findings recently suggested by $N$-body simulations \citep{Du+2017,Guo+2020}. 
Section \ref{sec:DISCUSSION} presents a discussion on the dependence of the CMO-to-galaxy stellar mass ratio on the bar-pattern speed. Finally, Section \ref{sec:Summary} concludes the paper.

\section{Methods and Models}
\label{sec:ModelMethod}

This section presents the gravitational potential of a double-barred galaxy. Further, the concept of loops and a method for determining the initial conditions required for a particle to form a loop are described.

\subsection{Galaxy model}
\label{sec:model:galaxy}

The galaxy model adopted in this study is based on Model 2 used in \citet{MaciejewskiSparke2000}. The components were a spheroid (i.e., stellar bulge + dark matter halo), a stellar disc, and two stellar bars; their respective density distributions are described below.
For the bulge and halo components, we adopted the following density distribution of the modified Hubble profile:
\begin{equation}
  \rho(r)=\rho_{\mathrm{b}}\left(1+\frac{r^{2}}{r_{\mathrm{b}}^{2}}\right)^{-1.5},
\label{eq:modifiedHubbleprofile}
\end{equation}
where $r^2=x^2+y^2+z^2$, $r_{\mathrm{b}}$ is the scale length of the bulge, and $\rho_{\mathrm{b}}$ is the central density.
The disc component adopted the surface density distribution of the Kuzmin--Toomre model.
\begin{equation}
  \Sigma(R)=\Sigma_{\rm d,0}\left(1+\frac{R^{2}}{R_{\rm d}^{2}}\right)^{-1.5},
\label{eq:Kuzmin-Toomreprofile}
\end{equation}
where $R^2=x^2+y^2$, $\Sigma_{\rm d,0}$ is the central surface density and $R_{\rm d}$ is the scale length of the disc. The potentials of both the density distributions can be calculated analytically \citep{BinneyTremaine2008}.

\begin{figure}
\begin{center}
\includegraphics[width=0.45\textwidth]{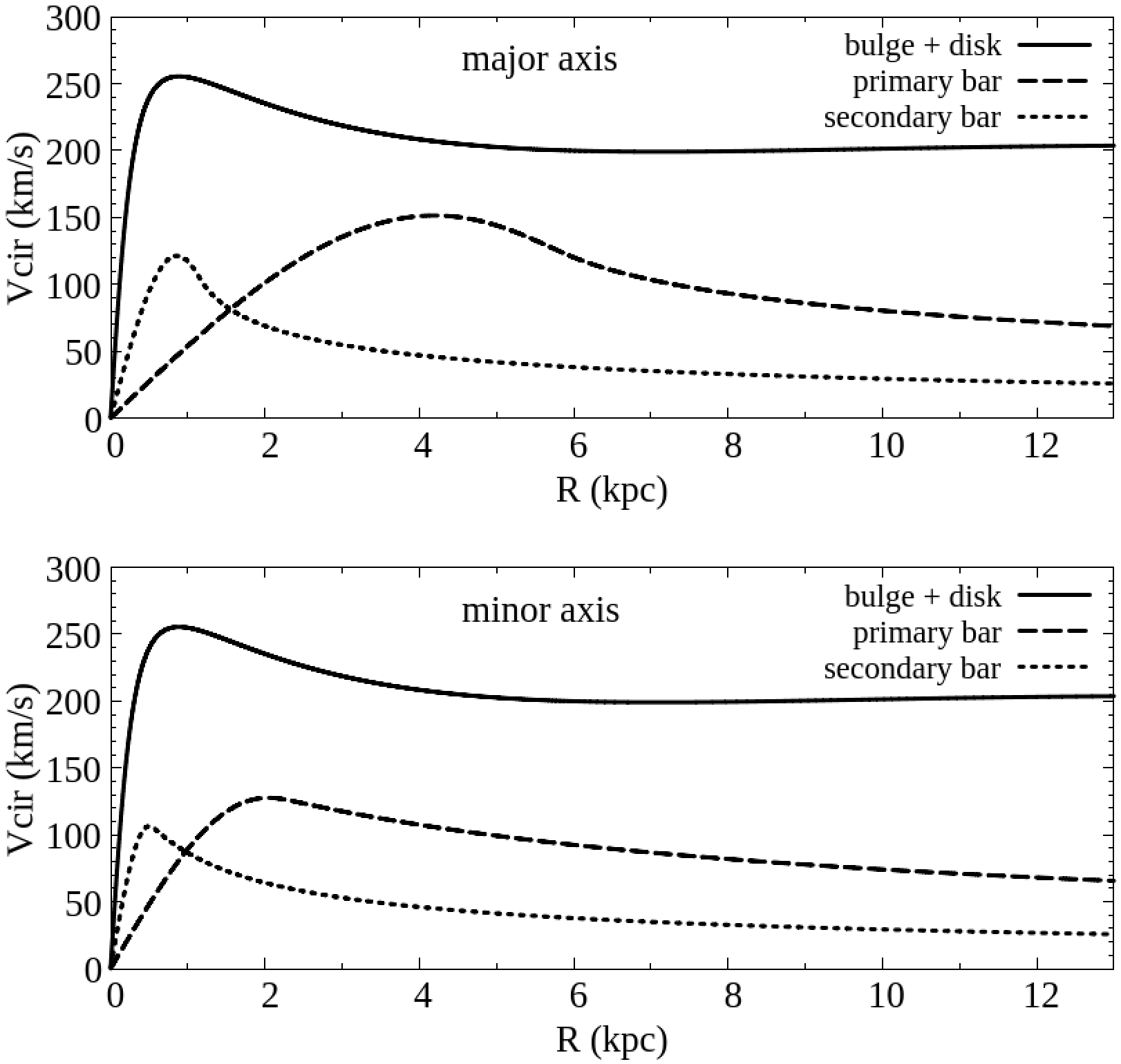}
\end{center}
 \caption{
 Circular velocity curves along the major axis (top) and minor axis (bottom) of the {primary} bar of the M0 model. The major axis of the {secondary} bar is aligned with the major axis of the {primary} bar.
 The solid, dashed, and dotted lines represent the axisymmetric component (bulge + disc), primary bar, and secondary bar, respectively.
 }
 \label{fig/vcir.png}
\end{figure}

For the two-bar components, the density distribution of the prolate Ferrers ellipsoid is assumed to be
\begin{equation}
  \rho(x, y, z)=
  \left\{
  \begin{array}{ll}
  \rho_{0}\left(1-m^{2}\right)^{n} & {\rm if}~~m<1, \\
  0 & {\rm  otherwise},
  \end{array}\right.
\label{eq:Ferrersellipsoid}
\end{equation}
where $m^{2}=x^{2} / a^{2}+y^{2} / b^{2}+z^{2} / c^{2}$ and $\rho_{0}$ represents the central density with $n=2$.
The gravitational potential and acceleration of the Ferrers ellipsoid cannot be determined analytically from the above density distribution. This study employed the polynomial expansion approximation of \citet{Pfenniger1984b} to calculate the gravitational potential and accelerations of $n=2$ Ferrers ellipsoids\footnote{The original equations in \citet{Pfenniger1984b} include certain typos. We fixed these by following \citet{OllePfenniger1998}.}.
The numerical values for each parameter are summarised in Table~\ref{tab:galaxy_Parameters}.
The circular velocity curve ($V_{\rm cir}$) is shown in Figure~\ref{fig/vcir.png}. 

\begin{figure}
\begin{center}
\includegraphics[width=0.45\textwidth]{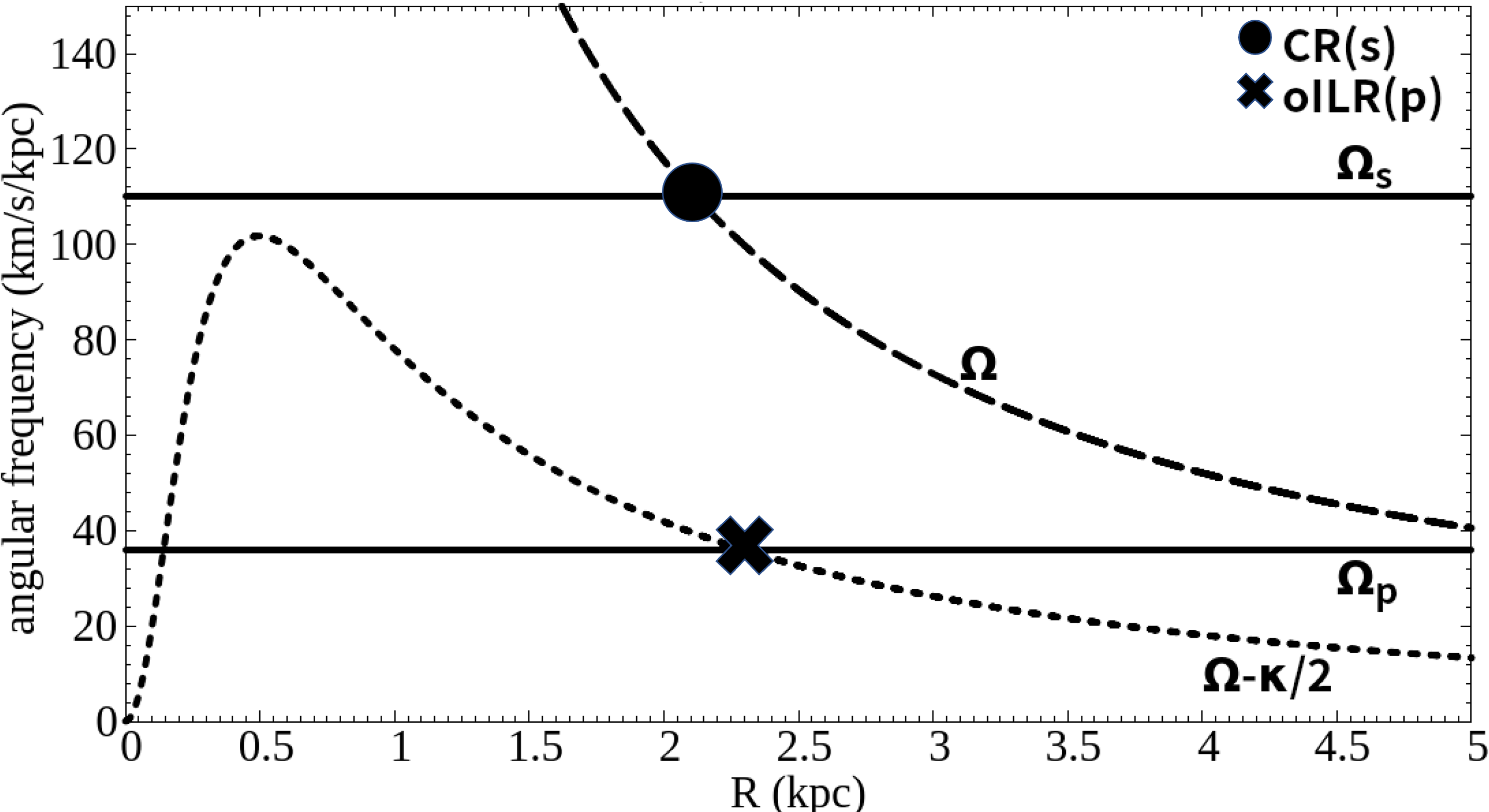}
\end{center}
  \caption{
  {Angular frequency curve of the inner region of the M0 model. The dashed and dotted lines indicate the angular frequency curves of $\Omega$ and $\Omega-\kappa/2$, respectively. The two horizontal lines indicate the pattern speeds of the primary ($\Omega_{\rm p}$) and secondary ($\Omega_{\rm s}$) bars.
  The black circle at approximately 2.2 kpc is the location of the corotation resonance of the secondary bar, CR(s).
  The cross at approximately 2.3 kpc represents the location of the outer ILR of the {primary} bar, oILR(p).
  }
  }
 \label{fig/longAngularFrequency.png}
\end{figure}

Figure~\ref{fig/longAngularFrequency.png} shows the angular velocity curves ($\Omega$; dashed line) and $\Omega-\kappa/2$ (dotted line) of the M0 model.
Here, $\Omega \equiv V_{\rm circ}/R$ and $\kappa$ are the angular and radial epicyclic frequencies, respectively, which are evaluated from the axisymmetric gravitational potential of the galaxy \citep{BinneyTremaine2008}. 
Further, $\Omega_{\rm p}$ and $\Omega_{\rm s}$ denote the pattern speeds of the primary and secondary bars, respectively. As evident, the primary bar has two inner Lindblad resonances (ILRs) at approximately 0.15 and 2.3 kpc, and a co-rotation (CR) resonance at approximately 2.2 kpc. Further, the secondary bar has no ILR but a CR at approximately 2.2 kpc. 
These resonance radii changed in the presence of CMC (see Section \ref{sec:model:cmc}).

\begin{table}
    \caption{
    Summary of parameters of the \citet{MaciejewskiSparke2000} galaxy model.
    In the Ferrers ellipsoid parameter $\rho_{0}$, $a$, $b$, and $c$, the `{p}' subscript corresponds to the {primary} bar and the `{s}' subscript corresponds to the {secondary} bar.
    }
    \centering
    \begin{tabular}{c|c}
      \hline 
      \ Disc parameters                                             \\
      \hline 
      $R_{\rm d}$        & 14.1~kpc \\
      $\Sigma_{\rm d,0}$    & 1.83~$\mathrm{M}_{\odot}~\mathrm{kpc}^{-2}$\\
      \hline
      \ Spheroid parameters                                    \\
      \hline
      \  $\rho_{\mathrm{b}}$
        &   4.6$\times 10^{10}~\mathrm{M}_{\odot}~\mathrm{kpc}^{-3}$  \\
      \  $r_{\mathrm{b}}$
        &   0.3~kpc                                                        \\

      \hline
      \ Primary bar parameters                                    \\
      \hline
      \  $\rho_{\mathrm{0,{p}}}$
        &  0.0986~$\mathrm{M}_{\odot}~\mathrm{pc}^{-3}$                     \\
      \  $a_{\mathrm{{p}}}$
        &   6~kpc                                                          \\
      \  $b_{\mathrm{{p}}}$
        &   2.28~kpc                                                       \\
      \  $c_{\mathrm{{p}}}$
        &   0.6~kpc                                                        \\
      \  $\Omega_{\mathrm{{p}}}$
        &   $36~\mathrm{~km} \mathrm{~s}^{-1} \mathrm{kpc}^{-1}$           \\
      \hline
      \ Secondary bar parameters                                    \\
      \hline
      \  $\rho_{\mathrm{0,{s}}}$
        &   0.274~$\mathrm{M}_{\odot}~\mathrm{pc}^{-3}$                     \\
      \  $a_{\mathrm{{s}}}$
        &   2.4~kpc                                                        \\
      \  $b_{\mathrm{{s}}}$
        &   1.2~kpc                                                        \\
      \  $c_{\mathrm{{s}}}$
        &   0.57~kpc                                                       \\
      \  $\Omega_{\mathrm{{s}}}$
        &   $110~\mathrm{~km} \mathrm{~s}^{-1} \mathrm{kpc}^{-1}$          \\
      \hline
    \end{tabular}
    \label{tab:galaxy_Parameters}
\end{table}

\subsection{Central Massive Objects}
\label{sec:model:cmc}

To investigate the dynamic effects of CMOs in orbits with double bars, we introduced a fixed CMO in a double-barred potential. For simplicity, following previous studies \citep{ShenSellwood2004,Du+2017}, we modeled the gravitational potential of the CMO as a Plummer sphere: 
\begin{equation}
  \Phi_{\mathrm{CMO}}(r)=-\frac{G M_{\mathrm{CMO}}}{\sqrt{r^{2}+\epsilon_{\mathrm{CMO}}^{2}}},
\end{equation}
where $M_{\rm CMO}$ and $\epsilon_{\rm CMO}$ are the mass and scale lengths of CMO, respectively.

The mass of the CMO is an interesting parameter in the interaction between CMO and the {secondary} bar from $N$-body simulations studied by \citet{Du+2017} and \citet{Guo+2020}.
In this study, we introduced the $M_{\rm CMO}=0$, $10^6M_{\odot}$, $10^7M_{\odot}$, and $10^8M_{\odot}$ cases as models, because we considered that CMO increased its mass because of the action of the secondary bar and other factors. Expressed as a ratio of the galaxy mass, the ratios are $M_{\rm CMO}$/$M_{\rm galaxy}\approx 0$, $10^{-5}$, $10^{-4}$, and $10^{-3}$.
Because $\epsilon_{\rm CMO}$ controls the compactness of the CMO, we used $\epsilon_{\rm CMO}=10$ pc to mimic a compact CMO such as SMBH and NSC.
The parameters of the CMOs used in this study are summarised in Table \ref{tab:CMO_Parameters}.

\begin{figure*}
\begin{center}
\includegraphics[width=0.9\textwidth]{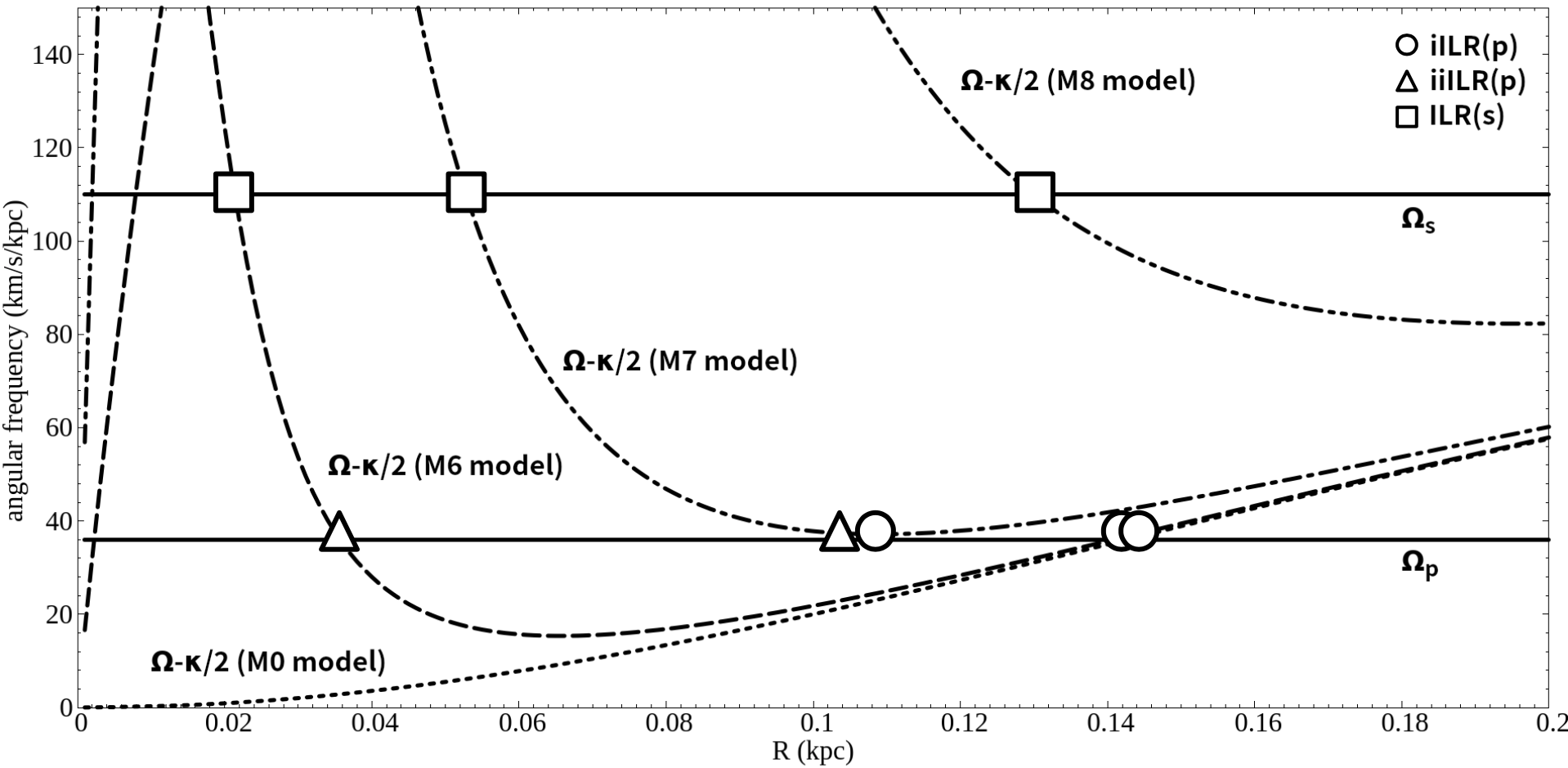}
\end{center}
  \caption{
  Transition of the angular frequency curves with the increasing mass of CMO.
  The upper and lower horizontal lines indicate the pattern speed of the {secondary} bar and {$\Omega_{\rm s}$}, and that of the {primary} bar and {$\Omega_{\rm p}$}, respectively.
  The dotted, dashed, single-dotted, and double-dotted lines represent $\Omega-\kappa/2$ in the M0, M6, M7, and M8 models, respectively.
  Open circles represent the inner ILR of the primary bar, iILR(p). 
  Triangles represent the inner ILR of the {primary} bar, iiILR(p). 
  Squares represent the ILR of the secondary bar, ILR(s).
  }
 \label{fig/AngularFrequency.png}
\end{figure*}

Figure~\ref{fig/AngularFrequency.png} shows $\Omega-\kappa/2$ as a function of $R$ for each model, with the horizontal lines representing $\Omega_{\rm p}$ and $\Omega_{\rm s}$. 
The intersection of these lines and curves indicated the radius of the ILR.
Open circles and triangles denote the locations of the inner ILR of the primary bar, iILR(p), and inner inner ILR of the primary bar, iiILR(p), respectively, whereas open squares indicate the location of the ILR of the secondary bar, ILR(s).
As evident, the increase in the mass of the CMO significantly affected the axisymmetric gravitational field of the galaxy, resulting in changes in $\Omega-\kappa/2$. Consequently, the location of the iILR(p) and the appearance of new resonances are shifted. 
The location of iILR(p) shifted inward and was situated at approximately $R = 0.1$ kpc in the M7 model, whereas it was located at approximately $R = 0.14$ kpc in the M0 and M6 models. The M8 model lacked the iILR(p) but identified two new resonances, the ILR(s) of the secondary bar and the iiILR(p) of the primary bar, in the M6 and M7 models. In the M7 model, the locations of iILR(p) and iiILR(p) were nearly identical, situated at approximately $R = 0.1$ kpc, and the location of iiILR(p) shifts outwards compared with the M6 model. Finally, in the M8 model, the ILR(s) moved further to approximately $R = 0.13$ kpc, and both iiILR(p) and iILR(p) disappeared.

\begin{table}
\caption{
Model names and corresponding CMO masses, along with the values normalized by galaxy mass. Note that the M0 model has no CMO.
}
\begin{center}
    \begin{tabular}{c|ccc}
      \hline 
      \  Model        &
      $\rm{M}_{\rm{CMO}}$       &
      $\rm{M}_{\rm{CMO}}$/$\rm{M}_{\rm{galaxy}}$
      \\
      \hline 
      \ M0               & $0~\rm{M}_{\odot}$   &   0   \\
      \ M6  & $10^6~\rm{M}_{\odot}$  &  $10^{-5}$          \\
      \ M7  & $10^7~\rm{M}_{\odot}$  &  $10^{-4}$          \\
      \ M8  & $10^8~\rm{M}_{\odot}$  &  $10^{-3}$          \\
      \hline
    \end{tabular}
\end{center}
\label{tab:CMO_Parameters}
\end{table}

\subsection{Multiperiodic orbits: loops}
\label{sec:Loop orbit}

This study focused on multiperiodic orbits to support the time-dependent gravitational potential of a two-dimensional (2D) double-barred galaxy, as invoked by \citet{MaciejewskiSparke1997,MaciejewskiSparke2000}. 
Before explaining these supporting orbits, we review orbital families in a rigidly rotating single bar.
A rigidly rotating single-bar potential comprises two main families of stable periodic orbits: $x_1$ and $x_2$ \citep{ContopoulosPapayannopoulos1980,Athanassoula+1983}.
In general, periodic orbits always move along the same curve (i.e., closed orbits). If they are stable, they form backbones of a steady potential; the nearby orbits are trapped around them \citep{BinneyTremaine2008}. Stable periodic orbitals are referred to as parent orbits. The stable periodic orbits in a single-bar potential, that is, the $x_1$ orbital family, comprise orbits elongated in the direction of the bar major axis, whereas the $x_2$ orbital family comprises orbits elongated perpendicular to the major axis. In the self-consistent single bar, most stars are in orbits trapped around the $x_1$ family \citep{SparkeSellwood1987,PfennigerFriedli1991}. Hence, the $x_1$ family constitutes the primary backbone of the orbital structure of a rigidly rotating single bar.

However, if a gravitational potential has a double bar, it pulsates with a $\omega_{\mathrm{p}}=2\left(\Omega_{\rm s}-\Omega_{\rm p}\right)$ frequency in a system where one bar is fixed (the factor of two is because of bisymmetry). Hence, in general, orbits in a double-barred potential lack a conserved integral of motion and are not closed in any rotating frame of reference. Instead, stars are often assumed to be chaotic, exploring large regions of the phase space. However, \citet{MaciejewskiSparke1997,MaciejewskiSparke2000} alluded that these potentials admit families of regular multiperiodic orbits. They postulated that these are closed curves, which, when populated with stars moving in a double-bar potential, return to their original positions every time the bars return to the same relative orientation. They referred to these curves as ``loops''. 
These are the double-frequency orbits driven by two bars \citep{MaciejewskiAthanassoula2007}.
Stars trapped around these stable loops could form the building blocks for a long-lived, double-barred galaxy \citep{MaciejewskiAthanassoula2008}, similar to the manner in which stars are trapped near the $x_1$ family in a single-barred galaxy. Hence, the loops are generalizations of the parent orbits in a single bar.

\begin{figure}
\begin{center}
 \includegraphics[width=0.45\textwidth]{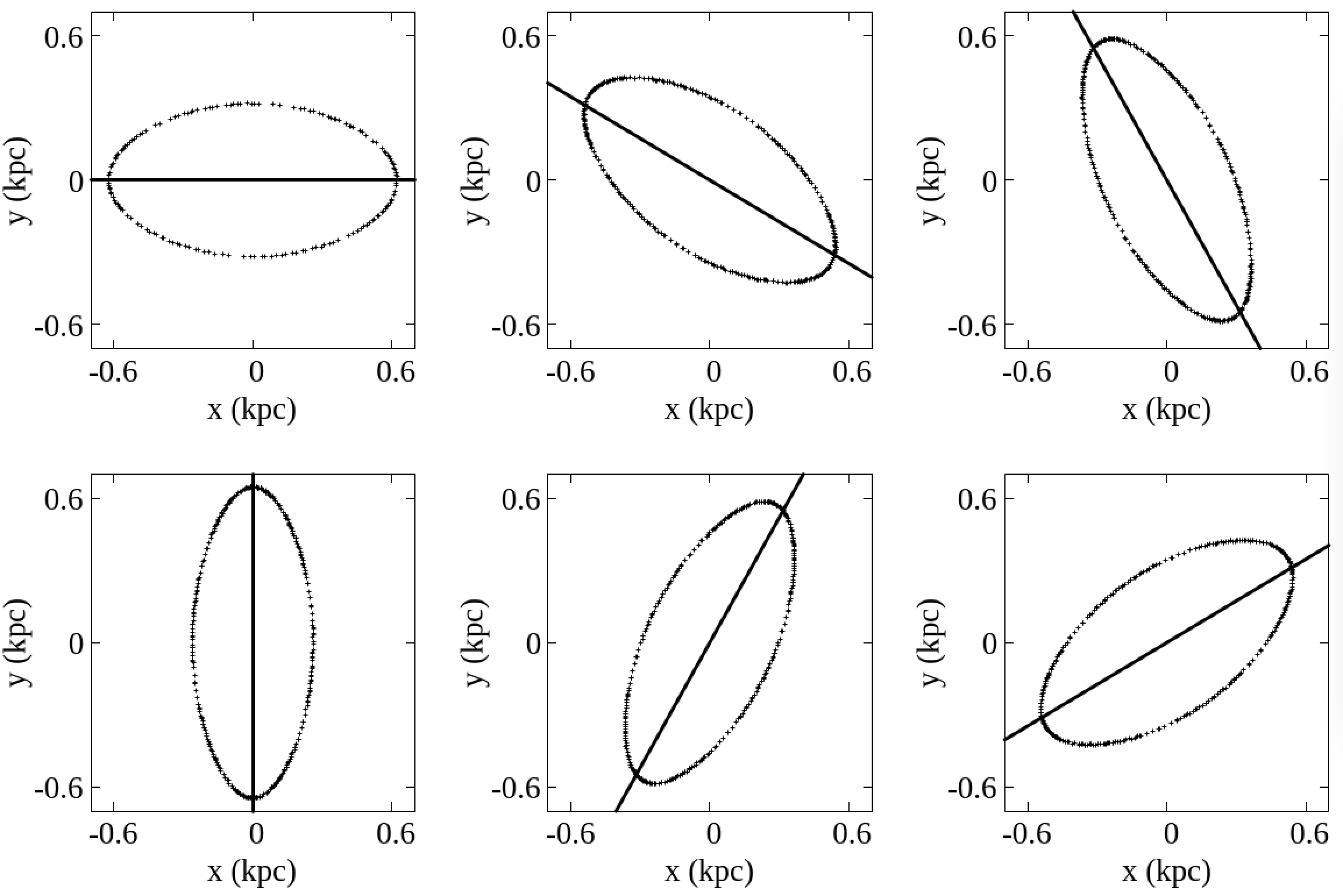}
\end{center}
 \caption{
 Example of a loop in the M0 model.
 The $x$-axis is the major axis direction of the {primary} bar, and the straight line is the major axis direction of the {secondary} bar.
 }
 \label{fig/LoopOrbit/LoopOrbit.png}
\end{figure}

Figure~\ref{fig/LoopOrbit/LoopOrbit.png} shows an example of a loop in the M0 model (i.e., no CMO) drawn by outputting stars when the relative direction of the bar becomes a specific angle. 
Based on the relative direction of the bar, the flatness of the closed orbit and alignment of the axis with the {secondary} bar are slightly different. However, the closed orbit rotates along the gravitational potential of the {secondary} bar. Thus, the loops shown in Figure~\ref{fig/LoopOrbit/LoopOrbit.png} can be considered the building block orbits of a self-consistent double-barred galaxy.

However, the self-consistency of loops is not always guaranteed. In fact, \citet{MaciejewskiSparke2000} suggested that a long-lived {secondary} bar may exist only when a corotation resonance (CR) of the {secondary} bar, $R_{\rm CR(s)}$ is located near the ILR of the primary bar, $R_{\rm ILR(p)}$. 
{This condition was satisfied in our galaxy models (Figure~\ref{fig/longAngularFrequency.png}).}

\subsection{Initial conditions and ring width}
\label{sec:ModelMethod/ring width diagram}

The ring width diagram method aims to determine the initial conditions required for a particle to form a `loop' orbit in a double-barred gravitational potential. Ring width ($w$) is the radial width of the ring drawn by the orbit when the orbit of a particle is calculated for a sufficiently long time in a galaxy. Here, $w$ of the loop orbit is very small, whereas that of an orbit with large free oscillation to the loop orbit is large.

To create the ring width map, we assumed that the major axes of the two bars coincided at $t = 0$ and the initial conditions for the particle were $(x, y, z, v_x, v_y, v_z) = (0, y_{\rm init}, 0, v_{x,\rm init}, 0, 0)$, where $y_{\rm init}$ and $v_{x,\rm init}$ are the initial values of $y$ and $v_x$, respectively. The equations of motion were integrated for 30 Gyr, and the positions and velocities of the particles were recorded only when the major axes of the primary and secondary bars were aligned using the loop property.

To obtain $w$, we divided the ring into 40 parts, wherein for each region, we considered the difference between the far and near points, denoted as $w_i~(i = 1, 2, ..., 40)$. Subsequently, the average value of $w_i$ was obtained as follows:
\begin{equation}
  w^{\prime}=\sum_{i=1}^{40} \frac{w_{i}}{40}.
\label{width_dash}
\end{equation}
Furthermore, $w'$ can be normalized by the total time average $r_{\rm ave}$ of the particle $r(t) = x^2 + y^2$. Consequently, the ring width $w$ can be obtained as
\begin{equation}
  w=\frac{w^{\prime}}{r_{\rm{ave}}}.
\label{width}
\end{equation}

By following this procedure, the initial conditions required for a particle to form a loop orbit with a double-barred gravitational potential can be determined.

\section{Results}
\label{sec:RESULTS}

\subsection{Ring width diagram and resonances}
\label{sec:RESULTS/ring width diagram}

\begin{figure*}
\begin{center}
\includegraphics[width=0.9\textwidth]{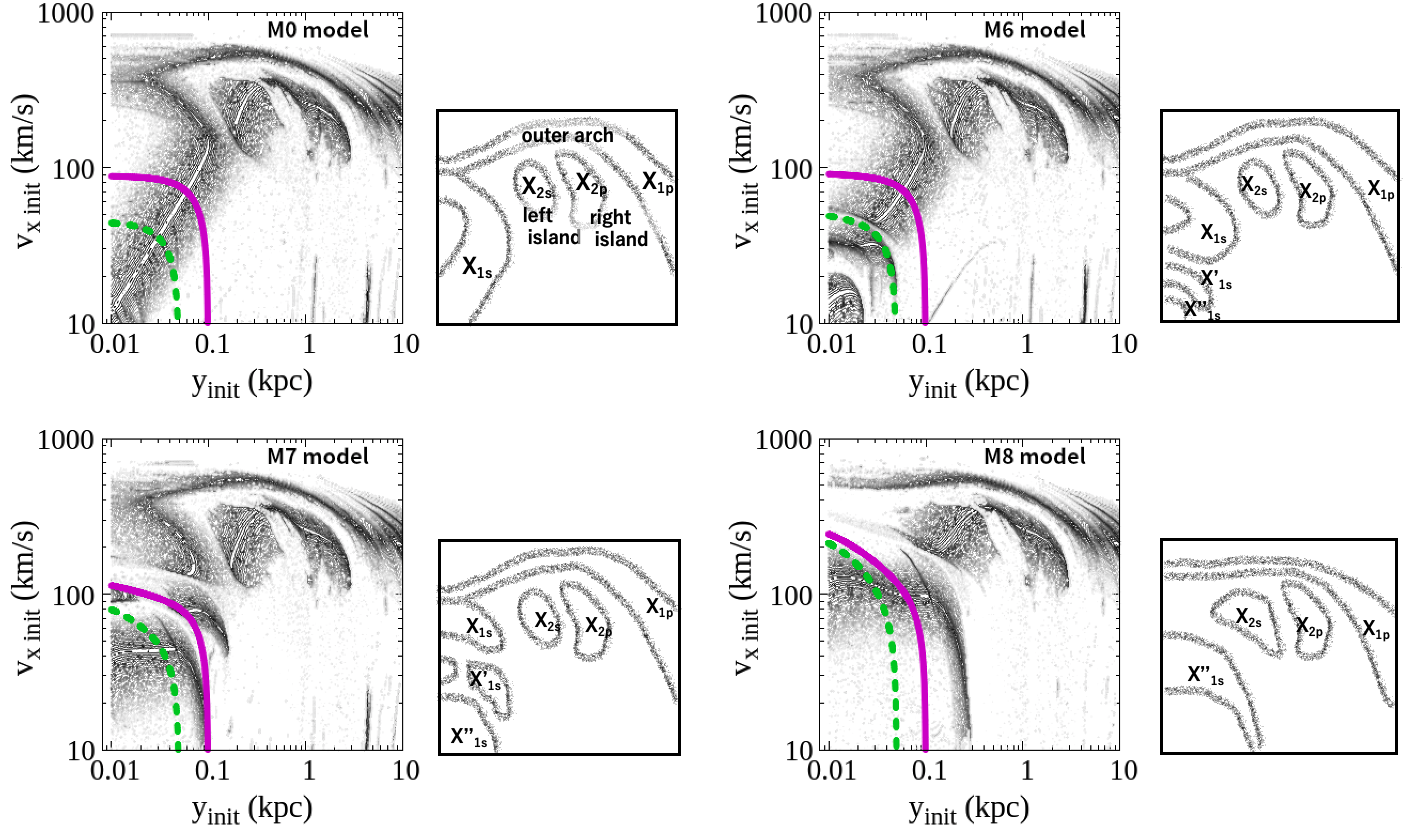}
\end{center}
\caption{
Ring width diagrams for models M0, M6, M7, and M8.
The horizontal and vertical axes denote $y_{\rm init}$ and $v_{\rm x,init}$, respectively. The gray shades indicate the ring width ($w$). The specific size of w in the trajectory derived from each initial condition is indicated through varying gray shades. Particles whose initial conditions are indicated by the black region are considered stable owing to their small ring width. Conversely, those indicated by the white region are considered unstable owing to their large ring width. The smaller the $w$, the more stable the initial conditions required to yield stable loops. Each black region is labeled with the corresponding loop name. The dotted green and solid purple lines indicate the iso-Jacobi curves for $y_{\rm max} = 0.05$ kpc and $y_{\rm max} = 0.1$ kpc, respectively (for more details on $y_{\rm max}$, refer to Section \ref{sec:RESULTS/Poincare map}). Schematics for each ring width diagram are shown on the right. The names of loop orbits corresponding to each region depicted in the ring width diagram are also displayed. 
}
\label{fig/RingWidthDiagram/10pc.png}
\end{figure*}

Figure~\ref{fig/RingWidthDiagram/10pc.png} shows an examination of the ring-width diagrams for each model, where the ring width ($w$) is represented by gray colors derived from the initial conditions determined by the horizontal and vertical axes. 
In the absence of a CMO, our M0 model replicated the analysis in \citet{MaciejewskiSparke2000}. The model exhibited a prominent outer arch and two inner islands, with the outer arch corresponding to the $x_1$ orbits of the primary bar and the two inner islands corresponding to the $x_2$ orbits of the primary bar.
However, for a double-bar system, the inner arch exhibits a gap, which divides it into left and right islands \citep{MaciejewskiSparke2000}, similar to our calculations. In this case, they showed that the right- and left-side islands comprised the $x_{\rm 2p}$ and $x_{\rm 2s}$ orbits, respectively. Therefore, the loops supporting the primary bar were the $x_{\rm 1p}$ and $x_{\rm 2p}$ orbits, and those supporting the secondary bar are the $x_{\rm 1s}$ and $x_{\rm 2s}$ orbits. 
This is because the major axes of $x_{\rm 1p}$ and $x_{\rm 2p}$ orbits and the major axis of the gravitational potential indicate that the primary bars are consistently nearly coincident. Similarly, the major axes of $x_{\rm 1s}$ and $x_{\rm 2s}$ orbits and the major axis of the gravitational potential indicate that the secondary bar is always almost coincident.

The introduction of a CMO into the model significantly altered the structure of the ring-width diagrams. The ring-width diagram of the M6 model is shown in the upper right panel of Figure~\ref{fig/RingWidthDiagram/10pc.png}. Two additional ``gaps'' were observed in the ring width distribution related to the $x_{\rm 1s}$ orbits owing to the presence of the CMO, with these new distributions tentatively named $x_{\rm 1s}'$ and $x_{\rm 1s}''$. 
With the increase in the CMO mass in the M7 and M8 models (lower left panels of Figure~\ref{fig/RingWidthDiagram/10pc.png}), these gaps shifted positions. Further, the $x_{\rm 1s}''$ orbital region expanded, whereas the gaps between the $x_{\rm 1s}$ and $x_{\rm 1s}'$ orbits and the $x_{\rm 1s}'$ orbits disappeared.

Through comparisons of the locations of the gaps with the resonance radii, the presence and growth of the CMO were found to substantially influence the orbital structures, particularly in the inner regions. 
In the M0 model, iILR(p) (open circle in Figure~\ref{fig/AngularFrequency.png}) corresponds to the break between the $x_{\rm 1s}$ and $x_{\rm 2s}$ orbits in the ring-width diagram (approximately $y_{\rm init}=0.1$--0.2, $v_{x,\rm init}=100$--300).
In the M6 model, ILR(s) and iiILR(p) (open squares and triangles in Figure~\ref{fig/AngularFrequency.png}, respectively) correspond to the two lower-left breaks in the ring-width diagram (approximately $y_{\rm init}=0.01$--0.1, $v_{x,\rm init}=30$--60 and approximately $y_{\rm init}=0.01$--0.04, $v_{x,\rm init}=10$--30, respectively).
In M7 model, the outward shifts of ILR(s) and iiILR(p) (squares and triangles in Figure~\ref{fig/AngularFrequency.png}, respectively) correspond to the movement of the two gaps to the upper right of the ring-width diagram (approximately $y_{\rm init}=0.01$--0.2, $v_{x,\rm init}=80$--200, and approximately $y_{\rm init}=0.01$--0.2, $v_{x,\rm init}=30$--90, respectively).
The changes in the M8 model were more intricate: the ILR(s) (open squares in Figure~\ref{fig/AngularFrequency.png}) corresponds to the break between the $x_{\rm 2s}$ and $x_{\rm 1s}''$ orbits in the ring-width diagram (approximately $y_{\rm init}=0.01$--0.4, $v_{x,\rm init}=100$--300). Further, the disappearance of iiILR(p) corresponds to the disappearance of the break in the ring-width diagram, and the $x_{\rm 2s}$ orbital in the ring-width diagram is deformed because iILR(p) vanishes as well. Thus,  the support on the left side (approximately $y_{\rm init}=0.05$--0.1, $v_{x,\rm init}=200$--400) is lost.
Hence, we conclude that the presence of a CMO and its growth significantly influence the orbital structures, particularly in the inner regions, owing to the alteration of the resonances and contribution to the emergence of gaps in the ring-width diagrams.

\subsection{Orbital properties}
\label{sec:RESULTS/Poincare map}

\begin{figure*}
\begin{center}
\includegraphics[width=0.9\textwidth]{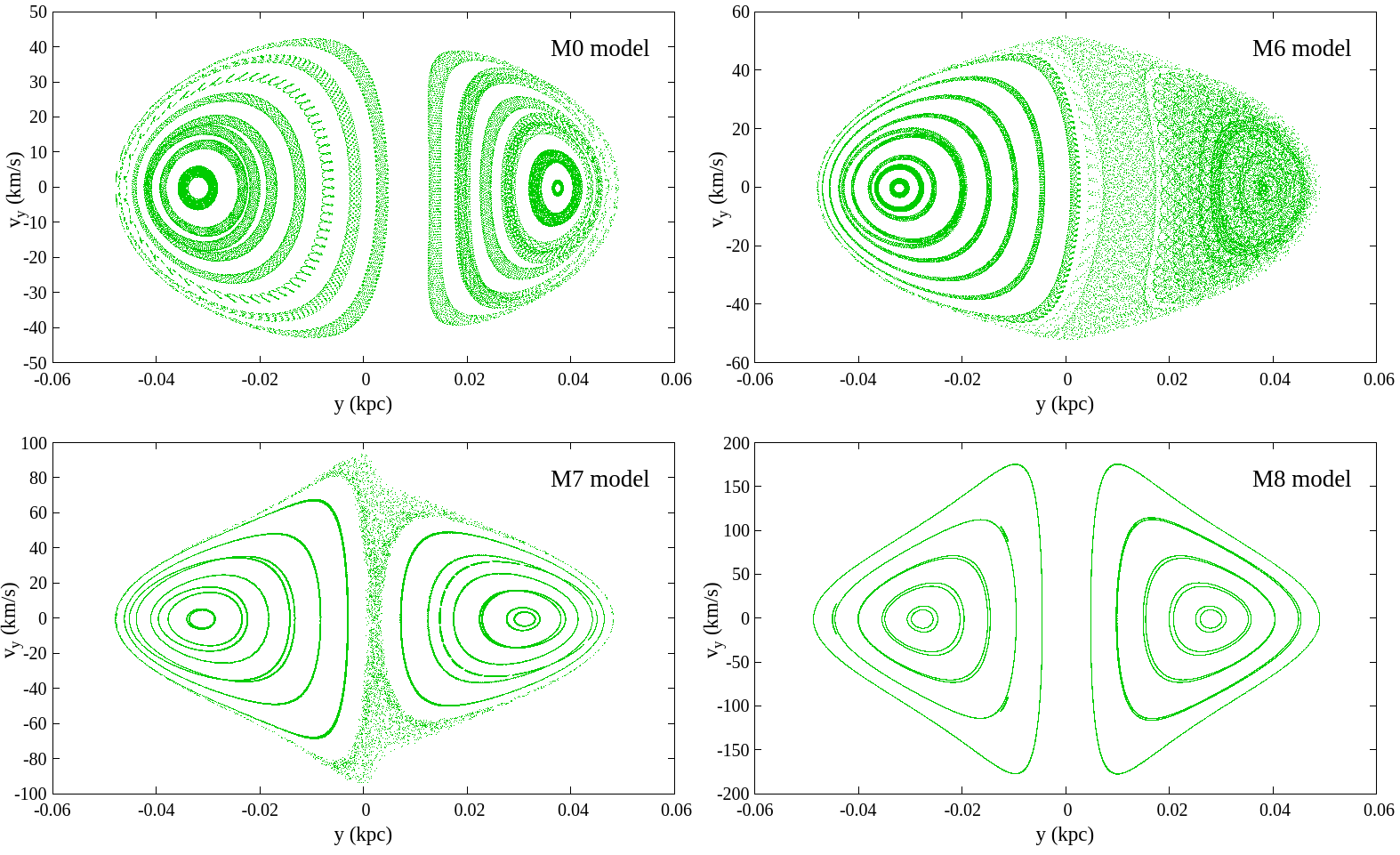}
\end{center}
  \caption{
  Poincare map created by the Jacobi integral of $y_{\rm{max}}=0.05~\rm{kpc}$.
  The corresponding Jacobi integral is indicated by the dotted green lines in Figure~\ref{fig/RingWidthDiagram/10pc.png}.
  }
 \label{fig/PoincareMap/50pc.png}
\end{figure*}

\begin{figure*}
\begin{center}
\includegraphics[width=0.9\textwidth]{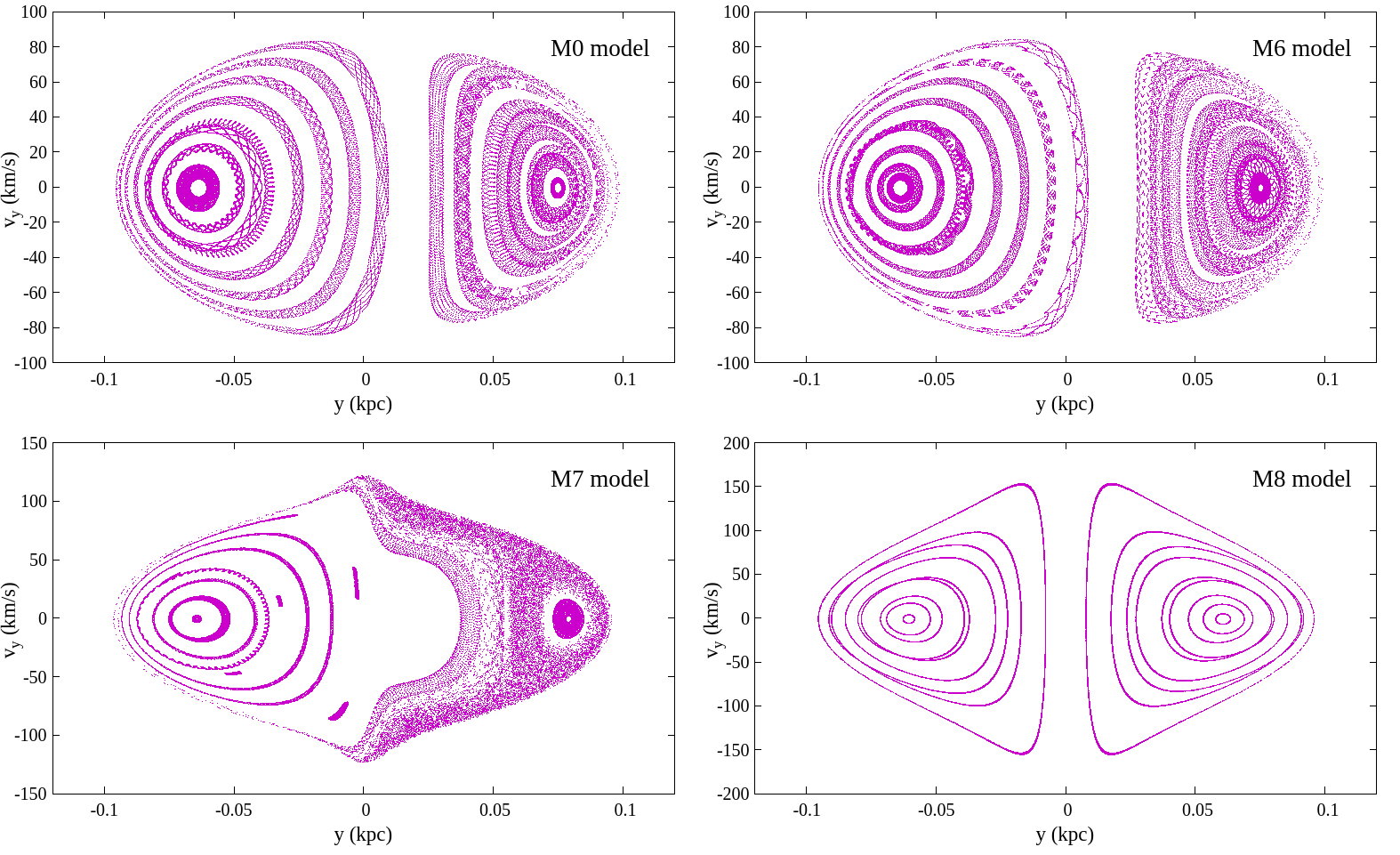}
\end{center}
  \caption{
  Same as Figure~\ref{fig/PoincareMap/50pc.png}, but for $y_{\rm{max}}=0.1~\rm{kpc}$. The corresponding Jacobi integral is indicated by the solid purple lines in Figure~\ref{fig/RingWidthDiagram/10pc.png}
  }
 \label{fig/PoincareMap/100pc.png}
\end{figure*}

As described in Section \ref{sec:RESULTS/ring width diagram}, we established the emergence of new orbital resonances in double-barred galaxies with a CMO, which resulted in significant dynamic shifts. To further investigate the effect of these resonances on the orbits, we analyze their properties, such as regularity or chaos, using the surface of section (SOS).

The SOS, a powerful tool for orbit analysis, is constructed at any accessible Jacobi energy \citep{BinneyTremaine2008,ShenSellwood2004}. Using Cartesian coordinates in the rotating frame of the secondary bar and aligning the major axis of the secondary bar with the $x$-axis, we recorded the values of $(y, v_y)$ for multiple orbits, each having the same $E_{\rm J,s}$ and crossing the minor axis ($y$-axis) as $v_x > 0$. In this frame, the Jacobi energy of a particle is expressed as $E_{\rm J,s} \equiv E - \Omega_{\rm s}L_z$, where $E$ and $L_z$ represent the energy and angular momentum around the $z$-axis, respectively.
The resulting $(y, v_y)$ SOS depicts the particles moving within a limited 2D space, where the maximum possible excursion ($y_{\rm max}$) on the minor axis of the secondary bar is determined by the given $E_{\rm J,s}$. 

Regular orbits that conserve another integral of motion in addition to $E_{\rm J,s}$, produce an invariant curve--a series of points lying on a closed curve in the SOS. By contrast, irregular chaotic orbits conserve only one integral of motion (i.e., $E_{\rm J,s}$) and occupy a region in the SOS.
Because the double-frequency orbits, which serve as parent orbits in double-barred galaxies, possess another conserved quantity (Section 2.3), they are distributed in a one-dimensional manner within the SOS. Our analysis enabled us to determine whether these orbit types are dominant at a given Jacobi energy.

Figs.~\ref{fig/PoincareMap/50pc.png} and \ref{fig/PoincareMap/100pc.png} show the SOSs with different $E_{\rm J,s}$, corresponding to $y_{\rm max} = 0.05$ and 0.1 kpc, respectively.
According to our definition, the right- and left-sides of the SOS are for forward and reverse rotation relative to the bar rotation, respectively.
Hereafter, we focus only on the right-hand sides of the SOSs because it is known that the $x_1$ and $x_2$ orbits (i.e., $x_{\rm 2p}$ and $x_{\rm 2s}$), which are important parent orbits of the bar, are in forward rotation with the bar.

Our SOS analyses revealed a distinct ``torus'' structure, representing one-dimensional invariant curves, in our M0, M7, and M8 models at a $y_{\rm max}$ of 0.05 kpc (Figure~\ref{fig/PoincareMap/50pc.png}). This structure is less evident in the M6 model, primarily because the iso-Jacobi energy curve (dotted green lines in Figure~\ref{fig/RingWidthDiagram/10pc.png}) intersects with gaps in the M6 model, as opposed to intersecting with the dark regions in the M0, M7, and M8 models. Similarly, a distinct ``torus'' structure can be observed in the M0, M6, and M8 models at $y_{\rm max} = 0.05$ kpc but not in the M7 model (Figure~\ref{fig/PoincareMap/100pc.png}). This indicates that the orbits within the loop were regular, whereas the gaps or resonances in the ring width diagram contained chaotic orbits.

In summary, these findings suggest that, with an increase in the CMO mass, the regular orbits are gradually disrupted from the center to the outer region owing to the shifting resonance location. Once the resonance passed, the CMO's spherically symmetric potential resulted in the formation of regular circular orbits that could not support the bar structure in a double-barred galaxy.

\subsection{Multiperiodic regular orbits}

\begin{figure*}
\begin{center}
\includegraphics[width=0.9\textwidth]{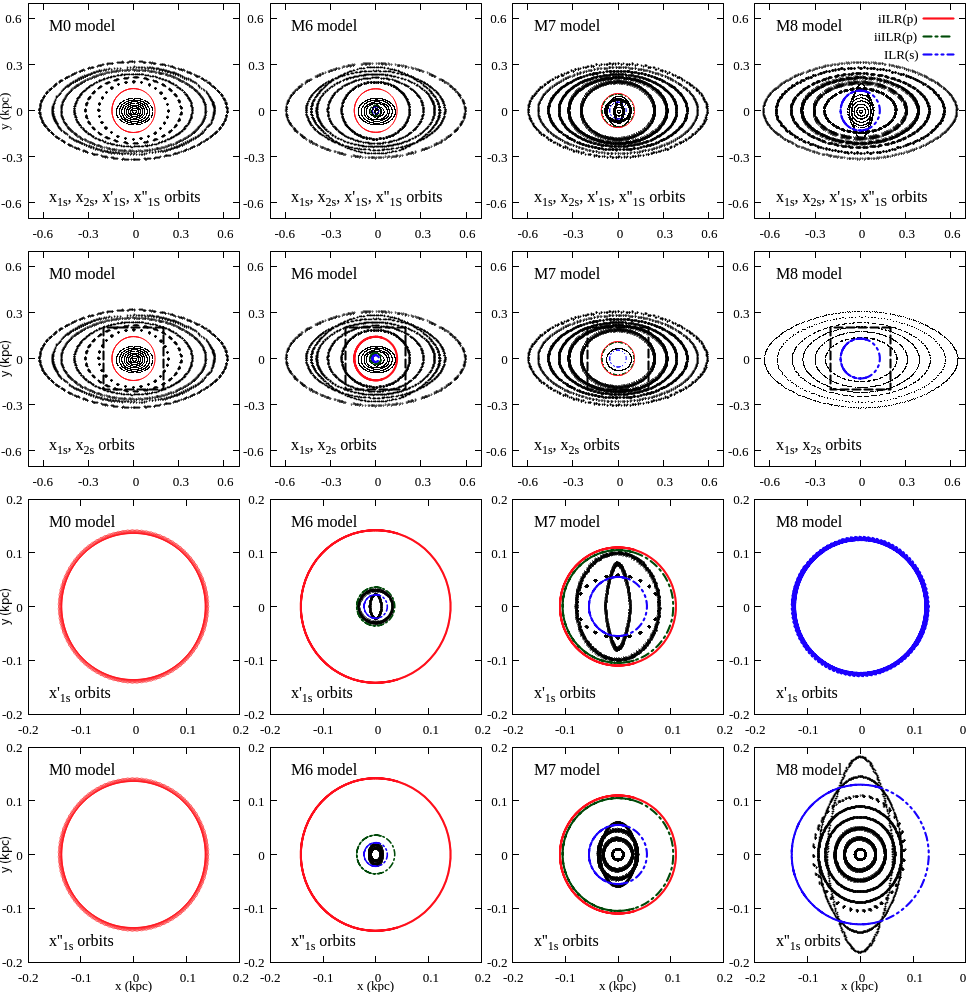}
\end{center}
\caption{
Examples of loops realized by each model. 
The figure illustrates the types of orbits that support the inner bar and are generated alongside the growth of the CMO. These orbits correspond to the $x_{\rm 1s}$, $x_{\rm 2s}$, $x_{\rm 1s}'$, and $x_{\rm 1s}''$ classifications.
The $x$-axis direction aligns with the major axes of the primary and secondary bars. The first to fourth columns display the orbits of the M0, M6, M7, and M8 models, respectively. The first-row depicts all the $x_{\rm 1s}$, $x_{\rm 2s}$, $x_{\rm 1s}'$, and $x_{\rm 1s}''$ orbits. The second-row illustrates only the $x_{\rm 1s}$ and $x_{\rm 2s}$ orbits, where in the M0, M6, and M7 models, the $x_{\rm 2s}$ orbits lie outside iILR(p) and the $x_{\rm 1s}$ orbits inside. In the M8 model, only the $x_{\rm 2s}$ orbit exists. The third row outlines only the $x_{\rm 1s}'$ orbit, which is non-existent in the M0 and M8 models. The fourth row sketches only the $x_{\rm 1s}''$ orbit, which is absent in the M0 model. The solid red, single-dotted green, and double-dotted blue lines indicate iILR(p), iiILR(p), and ILR(s), respectively.
}
\label{fig/LoopOrbit/LoopOrbits.png}
\end{figure*}

In this section, we analyze the multiperiodic regular orbits or loops of each model. Figure 8 shows the loops that uphold the secondary bars in each model. Figure~\ref{fig/LoopOrbit/LoopOrbits.png} focuses on the sub-kpc region at the center, where the major axis of the secondary bar is the $x$-axis. The resonance radii of the primary and secondary bars are marked with various line patterns. In all cases, a lack of an orbital family at these resonance locations and a transition to a different orbital family on either side were observed.

In the M0 model, the $x_{\rm 1s}$-orbital family is located inside the inner ILR of the primary bar iILR(p), whereas the $x_{\rm 2s}$-orbital family is located in the outer part. The size of the secondary bar is correlated with the area occupied by $x_{\rm 2s}$, where $x_{\rm 1s}$ occupies a significantly smaller area. Therefore, the $x_{\rm 2s}$-orbital family plays a significant role in maintaining a secondary bar.

As discussed in Section 3.1, new orbital families $x_{\rm 1s}'$ and $x_{\rm 1s}''$ emerged within the $x_{\rm 2s}$ orbital family in the models with a CMO. The $x_{\rm 1s}$ and $x_{\rm 2s}$ orbital families are presented in the second row of Figure~\ref{fig/LoopOrbit/LoopOrbits.png}, $x_{\rm 1s}'$ orbital family in the third row, and $x_{\rm 1s}''$ orbital family in the bottom row.
In the M6 and M7 models, variations in the orbital families related to the CMO occurred within the iILR(p), whereas the $x_{\rm 2s}$ orbital family remained relatively unaffected by the CMC. However, in the M8 model, the $x_{\rm 1s}''$ orbital family is sufficiently developed to overlap with the region occupied by the $x_{\rm 2s}$ orbital family.

We examined the inability of the secondary bar to gain support from the newly formed $x_{\rm 1s}''$-orbital family, which contributes to its destruction. The $x_{\rm 1s}''$ orbital family with its major axis perpendicular to that of the secondary bar does not align with the structure of the secondary bar. This is similar to the notion that the $x_2$ orbital family cannot provide consistent support for the bar in a single bar model \citep{BinneyTremaine2008}. Therefore, we expect that the $x_{\rm 1s}''$ orbital family will not be able to uphold a secondary bar as it becomes dominant. Briefly, the M8 model fails to maintain the secondary bar.

\section{Discussion}
\label{sec:DISCUSSION}

\subsection{Conditions for secondary bar destruction}
\label{sec:Discussion:condition}

\begin{figure*}
\begin{center}
\includegraphics[width=0.9\textwidth]{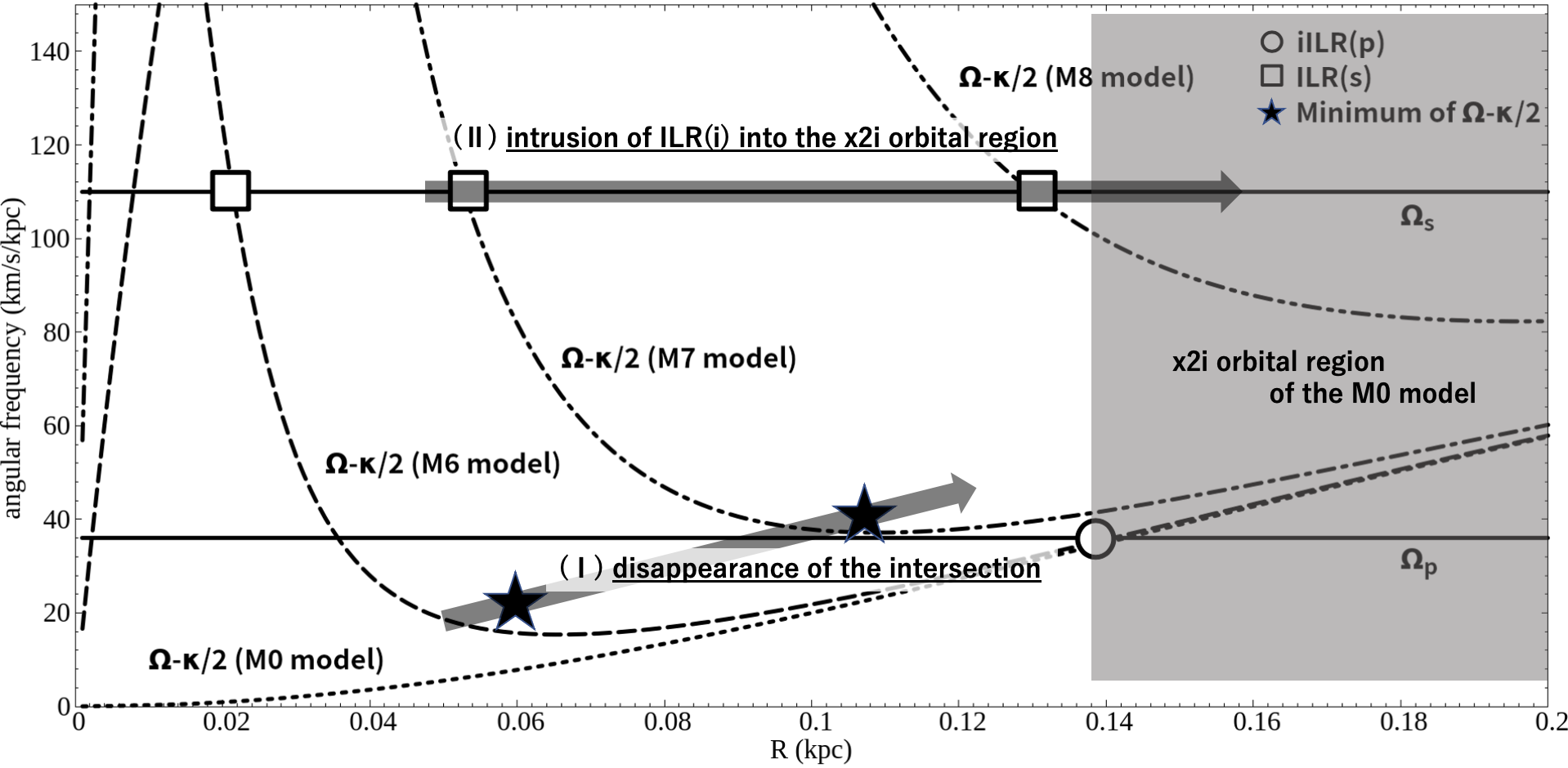}
\end{center}
  \caption{
  An enhanced version of Figure~\ref{fig/AngularFrequency.png} with additional explanations. The filled star symbols represent the minima of $\Omega - \kappa/2$, connected by an arrow. If the minima exceeds the $\Omega_{\rm p}$ line, it is labeled as [(I) disappearance of the intersection]. The gray-shaded region on the right side of the iILR(p) in the M0 model is referred to as the '$x_{\rm 2s}$ orbital region of the M0 model'. The intrusion of ILR(s) into this region, indicated by a square, is defined as (II) [intrusion of ILR(s) into the $x_{\rm 2s}$ orbital region].
  }
 \label{fig/Illustration of the discussion.png}
\end{figure*}

In this section, the conditions that lead to secondary bar destruction are explored. As discussed in the previous sections, it is postulated that the secondary bar disintegrates when the $x_{\rm 1s}''$ region, owing to its intrusion into the $x_{\rm 2s}$ region, loses self-consistency. Based on the results presented in Figure~\ref{fig/LoopOrbit/LoopOrbits.png}, in the M6--M7 models, where $x_{\rm 1s}''$ does not extend into the $x_{\rm 2s}$ region, the progression from the innermost region is $x_{\rm 1s}''$, ILR(s), iiILR(p), iILR(p), and $x_{\rm 2s}$. With the expansion of the CMO, the $x_{\rm 1s}''$, ILR(s), and iiILR(p) regions increase gradually. However, in the M7 model, the $x_{\rm 1s}''$ region remained small because of the minimal size of ILR(s). By contrast, in the M8 model, the disappearance of iiILR(p) and iILR(p) facilitated the expansion of ILR(s), resulting in $x_{\rm 1s}''$ orbits extending their region.

Conversely, a condition signifying that the secondary bar has already broken is the relocation of ILR(s) into the region previously occupied by $x_{\rm 2s}$ orbits. The radial expansion of ILR(s) prompts an enlargement of the $x_{\rm 1s}''$-orbit region, which already lost its self-consistency by the time it entered the $x_{\rm 2s}$ region. Consequently, the encroachment of ILR(s) into the $x_{\rm 2s}$ region is a strong indicator of secondary bar disintegration.

This scenario is illustrated in Figure~\ref{fig/Illustration of the discussion.png}. We label the first event (I) as the disappearance of the intersection and the second event (II) as the intrusion of the ILR into the $x_{\rm 2s}$ orbit region. Event (I) corresponds to the disappearance of iiILR(p) and iILR(p) when the minimum value of the rotation rate minus half the radial frequency ($\Omega - \kappa /2$) exceeded the pattern speed of the primary bar ($\Omega_{\rm p}$). Event (II) refers to the encroachment of ILR(s) and $x_{\rm 1s}''$ orbit into the $x_{\rm 2s}$ orbit region as ILR(s) moved beyond the gray zone in this figure. The conditions of the secondary bar disruption can be inferred from these two events, which occurred in conjunction with an increase in CMO mass, occurring between events (I) and (II).

We estimated the $M_{\rm CMO}/M_{\rm galaxy}$ ratio under the secondary bar destruction conditions. Given that Event (I) aligned closely with the M7 model and Event (II) approximated the M8 model, it is plausible to assume $M_{\rm CMO}$ at the time (I) is $10^7~\Modot$ and $M_{\rm CMO}$ at the time (II) is $10^8~\Modot$. In this scenario, the timing of the secondary bar disruption in our model aligns with a $M_{\rm CMO}/M_{\rm galaxy}$ ratio of approximately $10^{-3}$. This result is consistent with the findings of $N$-body simulations in previous studies \citep{Du+2017}.

\subsection{General conditions}

\begin{figure}
\begin{center}
\includegraphics[width=0.48\textwidth]{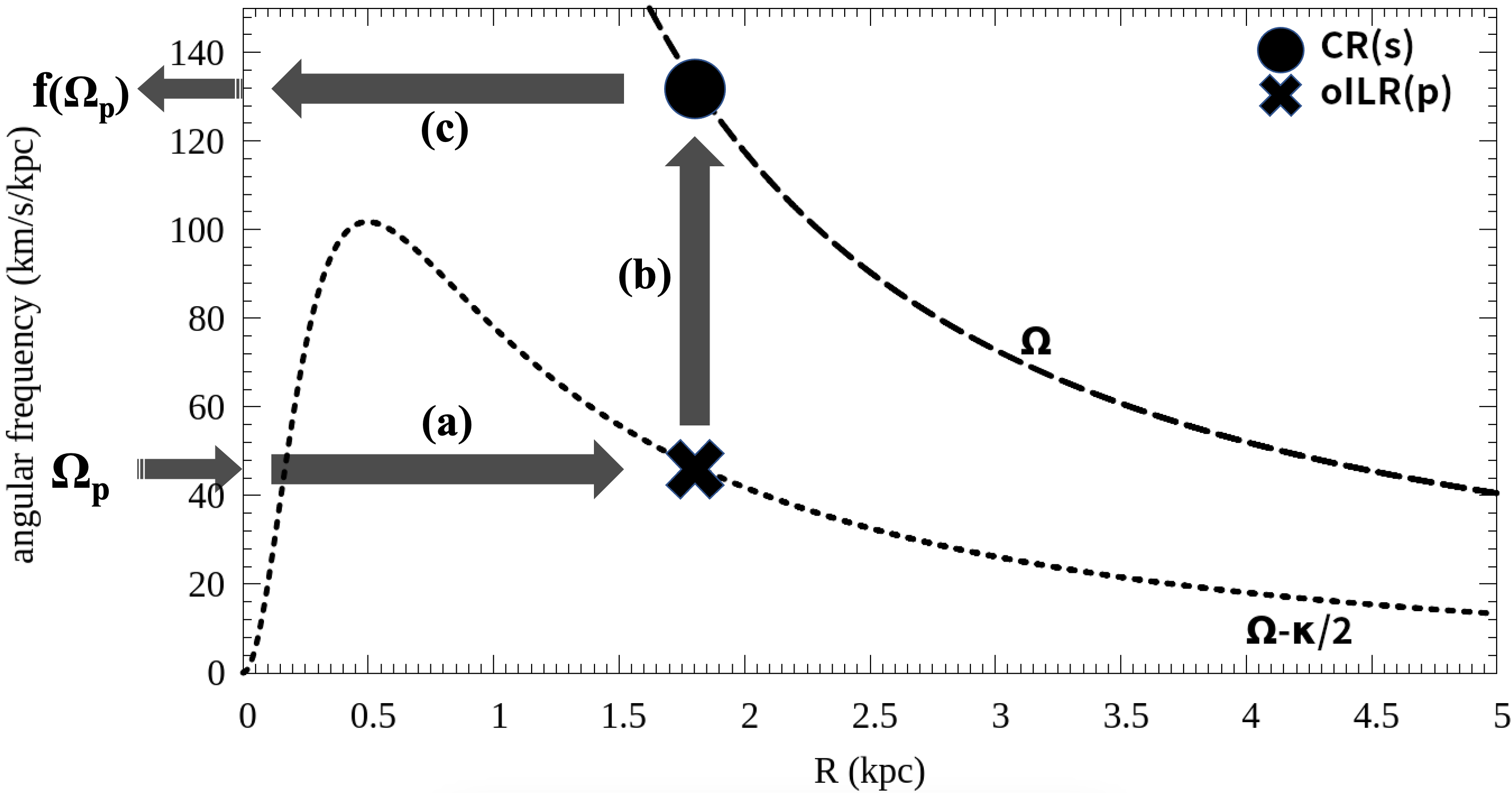}
\end{center}
  \caption{
  Schematic illustrating the relationship between $f(\Omega_{\rm p})$ derived from the determination of $\Omega_{\rm p}$. The procedures (a)-(c) correspond to eq.\ref{eq:Omega}: (a) Intersection with $\Omega-\kappa/2$ represents the outer ILR of the primary bar, $R_{\rm oILR(p)}$ (marked with a cross). (b) The CR radius of the secondary bar, $R_{\rm CR(s)}$ (marked with a filled circle), is obtained by imposing the condition $R_{\rm CR(s)} \approx R_{\rm oILR(p)}$. (c) The pattern speed of the secondary bar, $\Omega_{\rm s}$, can be determined from $R_{\rm CR(s)}$. 
  }
 \label{fig/F_omega.png}
\end{figure}

\begin{figure}
\begin{center}
\includegraphics[width=0.45\textwidth]{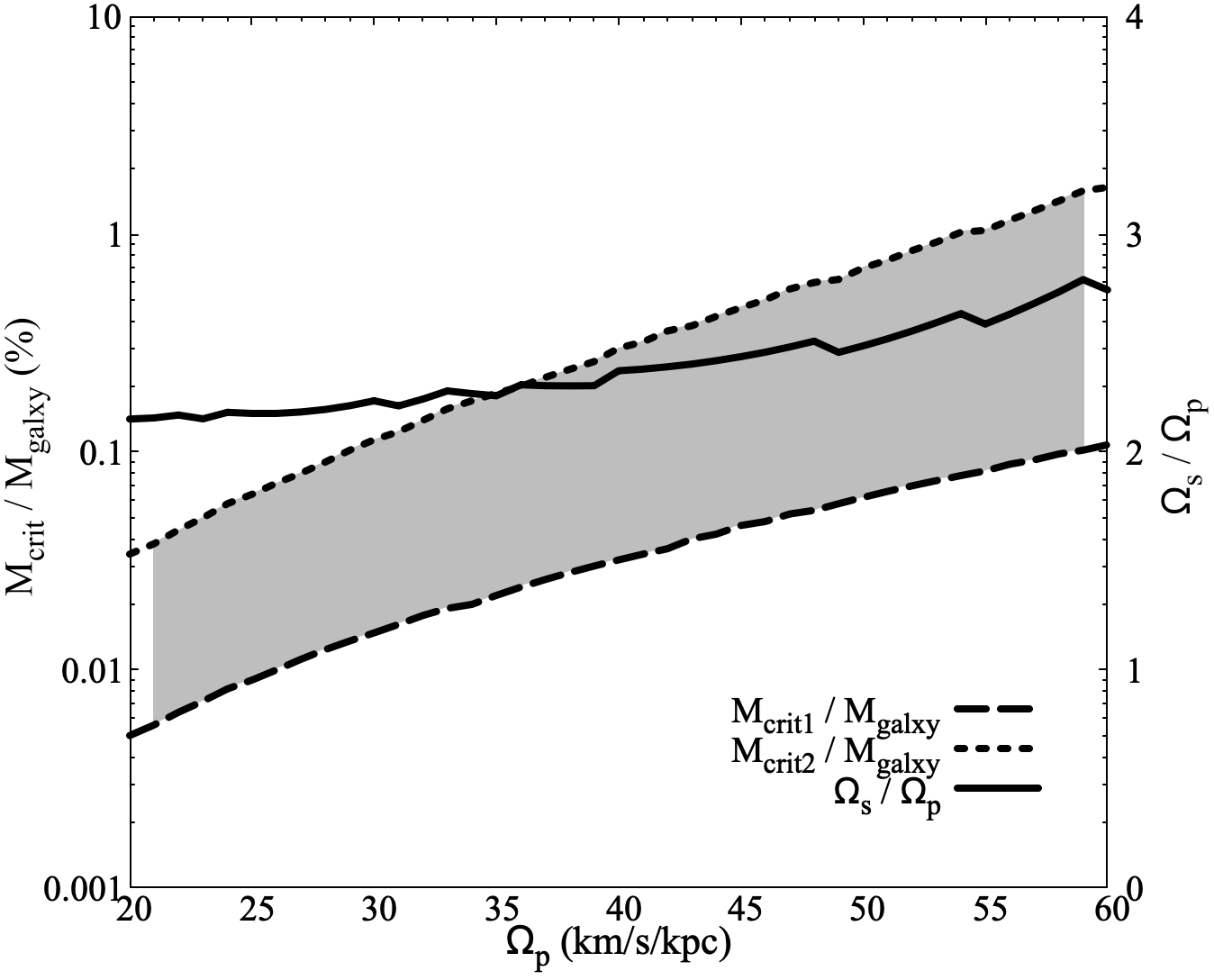}
\end{center}
\caption{
Critical CMO mass for secondary bar disruption.
The solid line indicated the ratio of secondary and primary bar pattern speeds ($\Omega_{\rm s}/\Omega_{\rm p}$) plotted against the primary bar pattern speed $\Omega_{\rm p}$, which is derived from Eq.(7).
The dashed line corresponds to the normalized critical CMO mass, $M_{\rm crit,I}/M_{\rm galaxy}$ at event (I), while the dotted line denotes the normalized critical CMO mass $M_{\rm crit,II}/M_{\rm galaxy}$ at event (II), both against $\Omega_{\rm p}$.
Events (I) and (II) refer to those defined in Section \ref{sec:Discussion:condition}. 
The gray area between (I) and (II) illustrates the range within which the secondary bar undergoes disruption.
}
\label{fig/Omega_vs_Mcrit.png}
\end{figure}

We now focus on the more general conditions. The preceding discussion relies on the results derived under the assumption of the pattern speeds applied in this study. However, as suggested by the discussion thus far, the effect of the CMO is pivotal in determining the position of the resonance radius, which is crucial for secondary bar disruptions. Consequently, we explored the dependence of the disruption conditions of the secondary bar on the bar pattern speed. Henceforth, we discuss the relationship between $M_{\rm CMO}/M_{\rm galaxy}$ and bar pattern speeds.

Given that the secondary bar forms within the radius of the outer ILR of the primary bar ($R_{\rm oILR(p)}$), the double bar can coexist stably if the co-rotation radius of the secondary bar ($R_{\rm CR(s)}$), which is approximately equivalent to the long radius of the secondary bar, and is within $R_{\rm oILR(p)}$. In this scenario, $R_{\rm CR(s)}\leq R_{\rm oILR(p)}$ holds true \citep{MaciejewskiAthanassoula2008,MaciejewskiSmall2010}. From the intersection of $\Omega - \kappa/2$ and $\Omega_{\rm p}$, $R_{\rm oILR(p)}$ is determined, and $\Omega_{\rm s}$ is set such that $R_{\rm CR(s)}$ is equal (see Figure~\ref{fig/F_omega.png}). If we denote this process by $f(x)$, the relationship is obtained as
\begin{equation}
    \Omega_{\rm s}=f(\Omega_{\rm p}).
\label{eq:Omega}
\end{equation}
The derived ratio $\Omega_{\rm s}/\Omega_{\rm p}=f(\Omega_{\rm p})/\Omega_{\rm p}$ is plotted in Figure~\ref{fig/Omega_vs_Mcrit.png}, yields a value in the range of 2.1--2.6, given that 20 \kmskpc{}$\lesssim\Omega_{\rm p}\lesssim$60 \kmskpc{} \citep{Aguerri+2003,Raitiainen+2008}.

For double-bar galaxies satisfying eq.(\ref{eq:Omega}), we calculated the CMO masses at which Events (I) and (II) occurred. The CMO masses where Events (I) and (II) occurred are denoted as $M_{\rm crit,I}$ and $M_{\rm crit,II}$, respectively. Figure~\ref{fig/Omega_vs_Mcrit.png} shows $M_{\rm crit,I}/M_{\rm galaxy}$ and $M_{\rm crit,II}/M_{\rm galaxy}$ for a given $\Omega_{\rm p}$. A notable feature is that with increase in $\Omega_{\rm p}$,  $M_{\rm crit,I}/M_{\rm galaxy}$ and $M_{\rm crit,II}/M_{\rm galaxy}$ increased slowly/blue{.}
As shown in Fig.~3, this is because the farther the line $\Omega_{\rm p}$ moves upward, the larger the CMO mass must be to eliminate the intersection of $\Omega-\kappa/2$ and $\Omega_{\rm p}$. Similarly, the more we move the line of $\Omega_{\rm p}$, the more the intersection point of $\Omega-\kappa/2$ and $\Omega_{\rm p}$ shifts to the left in the figure. Therefore, the mass of the CMO must be increased to slide the intersection point into the region of iILR(p).

As shown in Figure~\ref{fig/Omega_vs_Mcrit.png}, Event (I) occurred at $M_{\rm CMO}/M_{\rm galaxy}\approx10^{-4}$--$10^{-3}$ and Event (II) occurred at $M_{\rm CMO}/M_{\rm galaxy}\approx10^{-3}$--$10^{-2}$. As the secondary bar is projected to collapse between events (I) and (II), we can infer that a realistic double bar collapses when $M_{\rm CMO}/M_{\rm galaxy}$ ranges As $10^{-4}$--$10^{-2}$.
This aligns with the results of \citet{Du+2017} and \citet{Guo+2020}, who postulated that the secondary bar collapses when the CMO mass reaches approximately $10^{-3}$ of the total stellar mass. Thus, we can argue that this elucidates the mechanism of secondary bar destruction through an increase in CMO mass.

\section{Summary}
\label{sec:Summary}

This study investigated the stability of orbits supporting the double-bar configuration in galaxies by utilizing a Plummer sphere model to represent a CMO that has undergone mass growth owing to gas inflow. Previous studies have primarily focused on the stability of these orbits based on the pattern speed ratio between the primary and secondary bars. However, the results of this study revealed the significant influence of the CMO on double-bar orbit families, particularly when dividing the initial condition region responsible for generating loop orbits on the resonance map. 
The presence of the CMO altered the gravitational field in the central region, resulting in the emergence of ILRs associated with both primary and secondary bars. Consequently, certain double-bar orbit families could transition into unstable chaotic orbits. 

This highlights the novel finding that the existence of double-bar orbit groups is not solely determined by the relationship between the outer and secondary bars; rather, it is also influenced by factors such as the SMBH and other CMO effects. Additionally, the stability of the double-bar configuration is heavily reliant on the mass growth of the CMO driven by the secondary bar through the gas inflow. Our study provides a physical interpretation of the ``secondary bar--SMBH coevolution'' scenario based on previous $N$-body simulations, which encompasses the formation of the secondary bar, gas inflow, SMBH growth, and subsequent destruction of the secondary bar \citep{Du+2017,Guo+2020}.

Furthermore, we generalized the conditions for the destruction of the secondary bar. Based on our findings, we propose that a secondary bar is likely to disintegrate when $M_{\rm CMO}/M_{\rm galaxy}$ falls within the range $10^{-4}$--$10^{-2}$. This estimation provides valuable insights into the dynamics of double-barred galaxies, Further, it supports previous research conducted by \citet{Du+2017} and \citet{Guo+2020}, which suggested that augmentation of the CMO mass causes the disruption of double-bar structures. By considering the CMO effects, we contribute to the understanding of the secondary bar--SMBH coevolution scenario and shed light on the interplay between different components in galactic systems.

Regarding the implications for the Milky Way galaxy, the estimated range of destruction of the secondary bar ($M_{\rm CMO}/M_{\rm galaxy}\approx10^{-4}$--$10^{-2}$) is of particular interest. The Milky Way, with its primary pseudo-bulge formed through the secular evolution of a large-scale stellar bar \citep{Shen+2010}, exhibits a CMO (SMBH+NSC) mass of approximately 2--$4\times10^7$ \Msun{} \citep{Neumayer+2020}, which corresponds to approximately $\lesssim10^{-4}$ of the total stellar mass of the Milky Way. This value is comparable to or slightly smaller than the critical value suggested in our study and previous research \citep{Du+2017}. Therefore, the presence of a secondary bar in the Milky Way galaxy is uncertain; if it exists, it may be in a state of destruction. Observations of the number of red clump stars in the galactic bulge region have suggested the existence of a secondary bar \citep{Nishiyama+2006,Gonzalez+2011}; however, this evidence is still under debate \citep{GerhardMartinez-Valpuesta2012,Fujii+2019}. Thus, accurate measurements of the distances and velocities of stars in the Galactic nuclear bulge region are required to resolve this issue. Consequently, understanding the coevolution of the secondary bar and CMO based on stellar orbital motion, as demonstrated in our study, will be an important step in future observational studies on stars in the Galactic bulge.

Notably, our findings are based on test particle calculations and the utilization of a Plummer sphere model to represent the CMO. Further investigations and observational studies are required to validate and expand upon these results. Nevertheless, our research contributes to the growing body of knowledge on the dynamics and stability of double-bar configurations in galaxies.

\begin{ack}
JB acknowledges the support from JSPS Grant Numbers 18K03711, 21K03633, 21H00054, 18H01248, and 19H01933.
\end{ack}


\bibliographystyle{mn2e}
\bibliography{ms}

\end{document}